\documentclass[aps,prb,twocolumn,superscriptaddress,showpacs,floatfix,longbibliography]{revtex4-2}
\usepackage{graphicx,bm,epsfig,textcomp,color,dcolumn,setspace,array,mathrsfs,amsmath,amssymb,gensymb,booktabs,url,bibentry,natbib}
\usepackage{subfigure}
\usepackage{float}
\usepackage{bm}
\usepackage{braket}
\usepackage[mathscr]{eucal}
\usepackage{mathptmx}
\usepackage{xcolor}
\usepackage[bookmarks=true,colorlinks,linkcolor=blue,urlcolor=blue,citecolor=blue,
plainpages=false,pdfpagelabels,final,breaklinks=true]{hyperref}
\DeclareMathAlphabet{\mathcal}{OMS}{cmsy}{m}{n}

\begin{document}
\title{Voltage-driven exchange resonance achieving 100\% mechanical efficiency}
\author{Jun-Yu Tang}
\affiliation{Department of Physics and Astronomy, University of California, Riverside, California 92521, USA}
\author{Ran Cheng}
\affiliation{Department of Electrical and Computer Engineering, University of California, Riverside, California 92521, USA}
\affiliation{Department of Physics and Astronomy, University of California, Riverside, California 92521, USA}

\begin{abstract}
Magnetic resonances driven by current-induced torques are crucial tools to study magnetic materials but are very limited in frequency and mechanical efficiency. We propose an alternative mechanism, voltage-induced torque, to realize high efficiency in generating high-frequency magnetization dynamics. When a ferromagnet-topological insulator-ferromagnet trilayer heterostructure is operated as an adiabatic quantum motor, voltage-induced torque arises from the adiabatic motion of gapped topological electrons on the two interfaces and act oppositely on the two ferromagnetic layers, which can excite the exchange mode where the two ferromagnetic layers precess with a $\pi$-phase difference. The exchange mode resonance, bearing a much higher frequency than the ferromagnetic resonance, is accompanied by topological charge pumping, leading to a sharp peak in electrical admittance at the resonance point. Because the output current is purely adiabatic while dissipative current vanishes identically, the proposed voltage-driven exchange resonance entails a remarkably high mechanical efficiency close to unity, which is impossible in any current-driven systems.
\end{abstract}

\maketitle
\makeatletter 
\renewcommand{\@thesubfigure}{\hskip\subfiglabelskip}
\makeatother

\section{Introduction}\label{sec:Introduction}
Efficient manipulation of magnetization dynamics with low-energy dissipation has been a lasting inquiry in spintronics for decades~\cite{Hoffmann_Phys.Rev.Applied_2015,Baltz_RMP_2018,A.Manchon_RMP_2019}. Recently, there are growing interests in exploiting topological insulators (TIs) to achieve this goal~\cite{Kane_RMP_2010,S.C.Zhang_RMP_2011,X.L.Qi_PRB_2008}. For example, in a ferromagnet-topological (FM-TI) heterostructure, surface electrons are subject to spin-momentum locking, which can efficiently convert an applied charge current to a transverse spin polarization that exerts spin-orbit torques (SOTs) on adjacent magnetic moments~\cite{S.Akio_PRB_2014,Y.Wang_PRL_2015,A.Manchon_PRB_2017,Ghosh.S_PRB_2018,M.Z.Wu_SciAdv_2019,Y.Hyunsoo_NatComm_2017,D.C.Ralph_Nature_2014,D.C.Ralph_Nature_2014,K.L.Wang_NatMaterial_2014,Garate.Ion_PRL_2010,N.Naoto_PRB_2010}.

\begin{figure}[t]
\centering
\includegraphics[width=\linewidth]{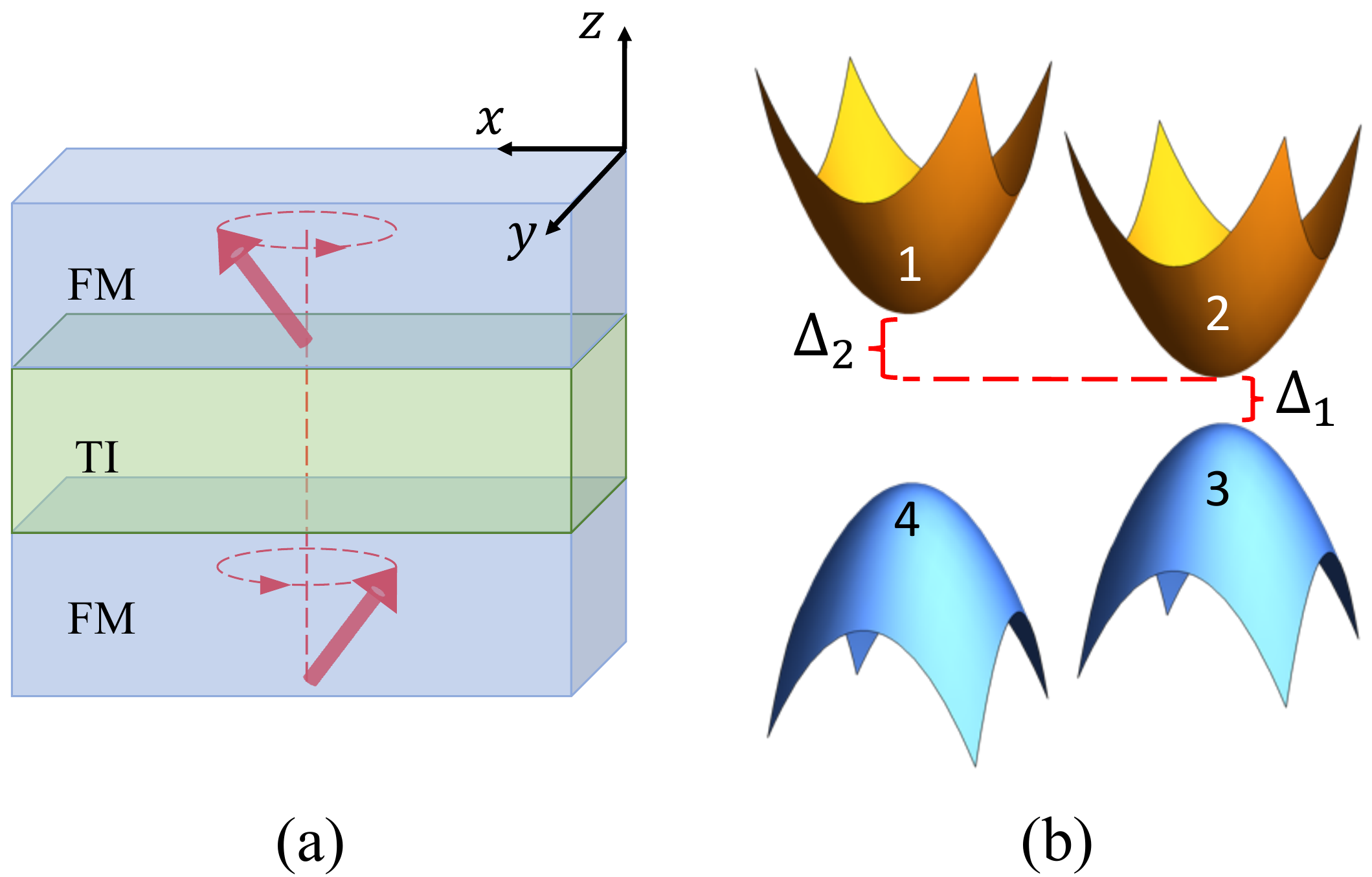}
\caption{(a) Illustration of the exchange mode in a FM-TI-FM trilayer and the coordinate system. (b) Band structure of the FM-TI-FM trilayer around the $\Gamma$ point, where $\Delta_1\sim J_{sd}-b_0$ and $\Delta_2\sim b_0$.}
\label{fig:model}
\end{figure}

While there are plenty of studies on current-induced magnetization switching~\cite{Y.Hyunsoo_NatComm_2017,D.C.Ralph_Nature_2014,K.L.Wang_NatMaterial_2014,M.Z.Wu_SciAdv_2019,C.Takahiro_PhysRevApplied_2020,J.P.Wang_NatMaterial_2018,L.Q.Liu_PRL_2017,STFMR_Liu,Yokoyama_PRB_2011,N.Naoto_PRB_2010,Garate.Ion_PRL_2010} and spin-torque ferromagnetic resonance~\cite{Y.Wang_PRL_2015,D.C.Ralph_Nature_2014,J.P.Wang_NatMaterial_2018,K.Kondou_NatPhys_2016} in FM-TI heterostructures, the operating speed is limited by the ferromagnetic spin dynamics---typically a few to a few tens of gigahertz. Recently, it has been proposed that in trilayer FM-TI-FM heterostructures, the interlayer exchange coupling mediated by topological electrons can be as strong as its counterpart in conventional spin valves with normal metal spacer~\cite{MingDa.Li_PRB_2015,B.A.Mansoor_AIPAdv_2017}. Such a trilayer system has two resonance modes: a ferromagnetic mode in which the two FM are in phase, and an exchange mode in which the two FM precess with a $\pi$-phase difference as illustrated by Fig.~\ref{fig:model}(a). The eigenfrequency of the latter, amplified by the interlayer exchange interaction, can be much higher than that of the former, opening a new pathway towards ultrafast spin dynamics driven by topological electrons.

However, leveraging the high-frequency exchange mode resonance (EMR) calls for driving forces of extremely high mechanical efficiency such that the concomitant waste heat will not damage the device. In traditional current-driven systems, the transport current inevitably incurs substantial Joule heating while only a small fraction of the input power eventually goes into magnetization dynamics. In TIs with surface states gapped by the proximity of magnetization, one usually needs to tune the Fermi level into the conduction band to retrieve a metallic interface so that an applied charge current can flow and generate non-equilibrium spin accumulation~\cite{Ghosh.S_PRB_2018,N.Naoto_PRB_2010,A.Manchon_PRB_2017,A.Manchon_RMP_2019,Yokoyama_PRB_2011}. Consequently, the conducting surface becomes dissipative, which significantly jeopardizes the mechanical efficiency. On the other hand, when the Fermi level lies in the gap (\textit{i.e.}, insulating surface), an applied voltage can generate the adiabatic motion of valence electrons devoid of dissipative currents~\cite{Xiao_PRB_2021,Garate.Ion_PRL_2010,S.Akio_PRB_2014,C.Takahiro_PhysRevApplied_2020,B.G.Xiong_PRB_2018}, which can then drive the magnetization as an adiabatic quantum motor (or Thouless motor) without Joule heating~\cite{Oppen_PRL_2013,arrachea2015nanomagnet}. When a FM-TI heterostructure is operated this way, the output current will be solely attributed to the magnetization precession (namely, topological charge pumping)~\cite{D.J.Thouless_PRB_1983}. As a result, the system barely consumes any energy when it remains stationary. In the vicinity of magnetic resonance, all input power goes into magnetization dynamics to overcome Gilbert damping, whereas Ohmic loss is eliminated. In such circumstances, the magnetization dynamics is actually driven by \textit{voltage-induced torques} rather than current-induced torques. However, it remains elusive whether voltage-induced torques are able to drive the desired high-frequency EMR in FM-TI-FM heterostructures.

In this paper, we first develop a semiclassical theory to quantify the voltage-induced SOT arising from the topological surface states of a TI exchange coupled to multiple insulating FMs, which, for Fermi level lying in the gap, is determined by the Berry curvature crossing the momentum and time spaces. Different from that in FM-TI bilayer systems, the effective magnetic field responsible for the SOT in a FM-TI-FM trilayer system has not only a parallel but also transverse components with respect to the direction of voltage drop. The parallel component, being the dominant contribution, is of opposite sign on opposite interfaces, which can directly excite the EMR when driving frequency matches the eigenfrequency. With the onset of EMR, the reciprocal effect of SOT, topological charge pumping, will generate a secondary adiabatic current on top of the driving force, forming a sharp peak of electrical admittance. Because the output current is purely adiabatic, which does not lead to Joule heating, all input power is cost by sustaining the magnetization dynamics, yielding a theoretical mechanical efficiency of $100\%$ unless the system is not perfectly insulating (\textit{i.e.}, in the presence of leakage current). The considered setting can be represented by an effective circuit with identical electrical responses. With all these unique properties combined, the voltage-driven EMR in trilayer FM-TI-FM heterostructures opens an exciting possibility to achieve high-frequency magnetization dynamics with almost zero dissipation.

The paper is structured as the following. In Sec.~\ref{sec:Formalism}, we formulate a Berry phase theory of the coupled dynamics of topological electrons and magnetization. In Sec.~\ref{sec:SOT}, we apply our formalism to an FM-TI-FM trilayer and numerically calculate the voltage-induced SOT. In Sec.~\ref{sec:EMR}, we study the voltage-driven EMR by calculating the dynamical susceptibility. In Sec.~\ref{sec:pumping}, we investigate the topological charge pumping effect associated with the EMR to quantify the overall electrical response of our system, and then discuss an experimental scheme to detect the EMR. In Sec.~\ref{sec:mecheff}, we evaluate the mechanical efficiency of the proposed setup and discuss the influence of imperfections (such as leakage current) our theory has ignored. The whole paper is summarized with concluding remarks in Sec.~\ref{sec:Summary}. Mathematical details about several derivations are clarified in the Appendices.

\section{Formalism}\label{sec:Formalism}

Let us consider free electrons moving in a multi-layer system where each magnetic layer is described by a uniform magnetization (macrospin model). In the presence of external electromagnetic field represented by a scalar potential $\varphi$ and a vector potential $\bm{A}$, the electronic Hamiltonian can be written as
\begin{align}
\label{eq:general Hamiltonian}
H=H_0(\hbar\bm{k}+e\bm{A},\bm{m}^1,\bm{m}^2,\cdots)-e \varphi,
\end{align}
where $H_0(\bm{A}\rightarrow0)$ is the unperturbed Hamiltonian of topological electrons on the surfaces of a TI, $e>0$ is the absolute value of the electron charge, $\hbar\bm{k}$ is the crystal momentum with $\hbar$ the reduced Planck constant, and $\bm{m}^j$ is the (dimensionless and unit) magnetization vector of the $j$-th magnetic layer. Here we allow $H_0$ to depend on all $\bm{m}^j$ because topological electrons on different interfaces are coupled, which will become clear in Sec.~\ref{sec:SOT} [see Eq.~\eqref{eq:Hamiltonian}] while here we keep the formalism general without specifying the form of $H_0$. Under the adiabatic regime where inter-band transition is negligible, an individual electron can be described by a semiclassical wave packet $|W(\bm{r}_c,\bm{k}_c)\rangle$ with $\bm{r}_c$ and $\bm{k}_c$ the center-of-mass position and momentum, respectively~\cite{Q.Niu_PRB_1999}. We then construct the Lagrangian for the wave packet as $\mathcal{L}_e=\langle W|(i\hbar d/dt-H)|W\rangle$~\cite{Cheng2013STT}. By adding the effective Lagrangian for each magnetic layer $\mathcal{L}_j=-N^j(\hbar S\phi^j\cos\theta^j+\varepsilon_j)$~\cite{G.Tatara_PhysicaE_2019} with $S$ the spin magnitude, $N^j$ the total number of magnetic atoms in layer $j$ (for cubic lattice, $N^j=N^j_xN^j_yN^j_z$), and $\varepsilon_j$ the magnetic free energy as a functional of $\bm{m}^j(\bm{r}_c, t)$ [and other $\bm{m}^i$ directly coupled to $\bm{m}^j$], we obtain the effective Lagrangian for an individual electron wave packet interacting with all magnetic layers as (see details in Appendix~\ref{sec:appendix EOM})
\begin{align}
\mathcal{L}=&\mathcal{L}_e + \sum_j\mathcal{L}_j=-
\bra{W}H\ket{W}-\sum_jN^j\varepsilon_j
\notag\\
&-\hbar\left(SN_j\dot{\phi}^j\cos\theta^j - \dot{\bm{r}}_c\cdot\bm{k}_c -\dot{\bm{m}}^j\cdot \bm{A}_{m^j} -\dot{\bm{k}}_c\cdot \bm{A}_k\right),
\label{eq:Lagrangian}
\end{align}
where the layer index $j$ is summed wherever repeated. All terms are evaluated at location $\bm{r}_c$. In Eq.~\eqref{eq:Lagrangian}, the $\dot{\phi}^j\cos\theta^j$ term accounts for the magnetization dynamics of layer $j$ where $\phi^{j}$ and $\theta^{j}$ are the two spherical angles specifying the orientation of $\bm{m}^j$; $\bm{A}_{m^j}=i\bra{u}(\partial/\partial\bm{m}^j)\ket{u}$ and $\bm{A}_k=i\bra{u}(\partial/\partial\bm{k})\ket{u}$ are the Berry connections in the magnetization and momentum spaces, respectively, where $\ket{u}$ is the periodic part of the Bloch wave function. By taking the functional derivatives of $\mathcal{L}$ with respect to $\bm{r}_c$, $\bm{k}_c$, and $\bm{m}^j$, we obtain a set of coupled equations of motion:
\begin{subequations}
\label{eq:eom}
\begin{align}
\hbar\bm{\dot{k}}_c &= -e\bm{E}\label{eq:EOMa},\\ 
\bm{\dot{r}}_c &=\frac{1}{\hbar}\frac{\partial\mathcal{E}}{\partial\bm{k}_c}+\dot{\bm{k}}_c\cdot\overleftrightarrow{\Omega}_{kk}+\dot{\bm{m}}^i\cdot\overleftrightarrow{\Omega}_{m^ik}\label{eq:EOMb},\\
N^jS\dot{\bm{m}}^j \times \bm{m}^j &=\frac{\partial\mathcal{E}}{\hbar\partial \bm{m}^j}+\dot{\bm{m}}^i\cdot\overleftrightarrow{\Omega}_{m^i m^j}+\dot{\bm{k}}_c\cdot\overleftrightarrow{\Omega}_{km^j}\label{eq:EOMc},  
\end{align}
\end{subequations}
where $i$ is summed while $j$ is a dangling variable specifying the magnetic layer under consideration, $\mathcal{E}=\bra{W}H\ket{W}+\sum_jN^j\varepsilon_j$, $\bm{E}=-\bm{\nabla}_c\varphi-\partial\bm{A}/\partial t$ is the applied electric field at center of the wave packet $\bm{r}_c$, and $\overleftrightarrow{\Omega}_{\mu\nu}=\partial_\mu \bm{A}_\nu-\partial_\nu\bm{A}_\mu$ is the Berry curvature tensor with $\mu$, $\nu$ running through each component of $\bm{k}_c$ and $\bm{m}^j$. The inner product of the tensor with a vector contracts the first sub-index while the second sub-index is dangling. For example, the $\nu$-component of $\dot{\bm{k}}_c\cdot\overleftrightarrow{\Omega}_{kk}$ refers to $\dot{k}_{c,\mu}\Omega_{k_{\mu}k_{\nu}}$ where $\mu$ is summed while $\nu$ specifies the spatial component of the resulting vector.

Equation~\eqref{eq:eom} is purely general as multiple physical effects are captured by different components of the Berry curvature tensor. The $\overleftrightarrow{\Omega}_{kk}$ term is the anomalous velocity responsible for the quantum anomalous Hall conductivity. The real-space component $\overleftrightarrow{\Omega}_{rr}=0$ because we do not consider inhomogeneous magnetization texture. The cross term $\overleftrightarrow{\Omega}_{m^ik}$ in Eq.~\eqref{eq:EOMb} is the Berry curvature connecting momentum $\bm{k}$ to the magnetization $\bm{m}_i$, hence, time $t$ [as $\bm{m}_i=\bm{m}_i(t)$]. It characterizes how magnetization dynamics affects the electron motion, namely, topological charge pumping. Reciprocally, the $\overleftrightarrow{\Omega}_{km^j}$ term in Eq.~\eqref{eq:EOMc} accounts for the back-action of the electron motion on the magnetization dynamics, which refers to the voltage-induced SOT. Finally, the $\overleftrightarrow{\Omega}_{m^im^j}$ term couples the magnetization of layer $i$ with that of layer $j$ through topological surface electrons; the $i=j$ component renormalizes the gyromagnetic ratio in a specific layer which will become clear in the following.

To derive the voltage-driven magnetization dynamics, we insert Eq.~\eqref{eq:EOMa} into Eq.~\eqref{eq:EOMc} to eliminate $\dot{\bm{k}}_c$.
To account for all Bloch electrons, we also integrate the electron-related terms over $\bm{k}_c$ within the first Brillouin zone (BZ) and $\bm{r}_c$ over the area of the interface: $\int d\bm{r}_c\int_{BZ}\mathcal{L}_ef(\bm{k}_c)d\bm{k}_c$, where $f(\bm{k})$ is the Fermi–Dirac distribution. For simplicity, we restrict to the low-temperature regime such that $f(\bm{k})=1$ for $\epsilon(\bm{k})<\epsilon_F$ and $f(\bm{k})=0$ otherwise. By taking the cross product $\bm{m}_j\times$ on both sides of the equation, we obtain the effective magnetization dynamics as (see details in Appendix~\ref{sec:appendix EOM})
\begin{align}
\dot{\bm{m}}^j=-\gamma\bm{m}^j\times(\overline{\bm{H}}^m_j+\bm{H}^{\rm SOT}_j- \frac1{N^j_zS}\sum_{i\neq j}\dot{\bm{m}}^i\cdot\overline{\Omega}_{m^i m^j}),
\label{eq:effectiveLLG}
\end{align}
where
\begin{align}
 \overline{\bm{H}}_j^m=-\frac{a_0^2}{(2\pi)^2 \hbar S}\int_{\rm BZ}\frac{\partial\varepsilon_j}{\partial\bm{m}^j}d\bm{k}_c \label{eq:Hj}
\end{align}
is the effective magnetic field acting on $\bm{m}^j$ in the absence of electric field with $N^j_z$ denoting the number of atomic layers in layer $j$ (note that $N^j=N^j_xN^j_yN^j_z$), and
\begin{align}
 \bm{H}_{j}^{\rm SOT}\equiv\frac{e\bm{E}}{N^j_z \hbar S}\cdot \overline{\Omega}_{km^j} \label{eq:SOT}
\end{align}
is the effective field for the voltage-induced SOT exerting on $\bm{m}^j$ with the overline representing an integration over the first BZ: $\overline{\Omega}_{km^j}\equiv a_0^2/(2\pi)^2 \int_{\rm BZ}\Omega_{km^j}d\bm{k}_c$. In deriving Eqs.~\eqref{eq:Hj} and~\eqref{eq:SOT}, we assumed that the whole system (TI and all magnetic layers) has a simple cubic lattice with lattice constant $a_0$, which does not sacrifice any essential physics. Since both $\bm{m}^j$ and $\gamma$ are dimensionless in our scaling convention, $\bm{H}^{{\rm SOT}}$ is expressed in the frequency dimension.

The last term of Eq.~\eqref{eq:effectiveLLG}, as mentioned above, represents the interlayer coupling mediated by topological electrons, which turns out to be a dissipative coupling~\cite{E.W.Gerrit_PRL_2003}. Compared with the interlayer Heisenberg exchange interaction~\cite{MingDa.Li_PRB_2015,B.A.Mansoor_AIPAdv_2017}, this dissipative coupling is negligible as estimated in Appendix~\ref{sec:appendix dissipative coupling}. In our convention, the bare gyromagnetic ratio is scaled to unity, while in Eq.~\eqref{eq:effectiveLLG},
\begin{align}
 \gamma\equiv\left[1+\frac{1}{N^j_z S}\bm{m}^j\cdot\overline{\bm{\Omega}}\right]^{-1} \label{eq:gamma_renorm}
\end{align}
is the effective gyromagnetic ratio~\cite{B.G.Xiong_PRB_2018} renormalized by the intra-layer component of the Berry curvature
\begin{align}
 \overline{\Omega}_{\lambda}\equiv\frac{1}{2}\epsilon_{\lambda\mu\nu}\frac{a_0^2}{(2\pi)^2}\int_{\rm BZ}d\bm{k}_c \Omega_{m^j_\mu m^j_\nu},\label{eq:vectorBerry}
\end{align}
where $\epsilon_{\lambda\mu\nu}$ is the Levi-Civita symbol and $\Omega_{m^j_\mu m^j_\nu}$ refers to the $\mu\nu$ element of the Berry curvature tensor $\overleftrightarrow{\Omega}_{m^j m^j}$ lying in the configurational phase space (\textit{i.e.}, magnetization direction). An intuitive understanding of the renormalization of $\gamma$ is that the topological electrons are polarized by the local magnetization, thus contributing partially to the total magnetic moment. By numerically evaluating the intra-layer components of $\overleftrightarrow{\Omega}_{m^j m^j}$, we estimate that the renormalization is only $10^{-5}$, so in the following we will use $\gamma=1$ for simplicity.

\section{Voltage-induced SOT in FM-TI-FM trilayer}\label{sec:SOT}

\begin{figure*}[t]
\includegraphics[width=\linewidth]{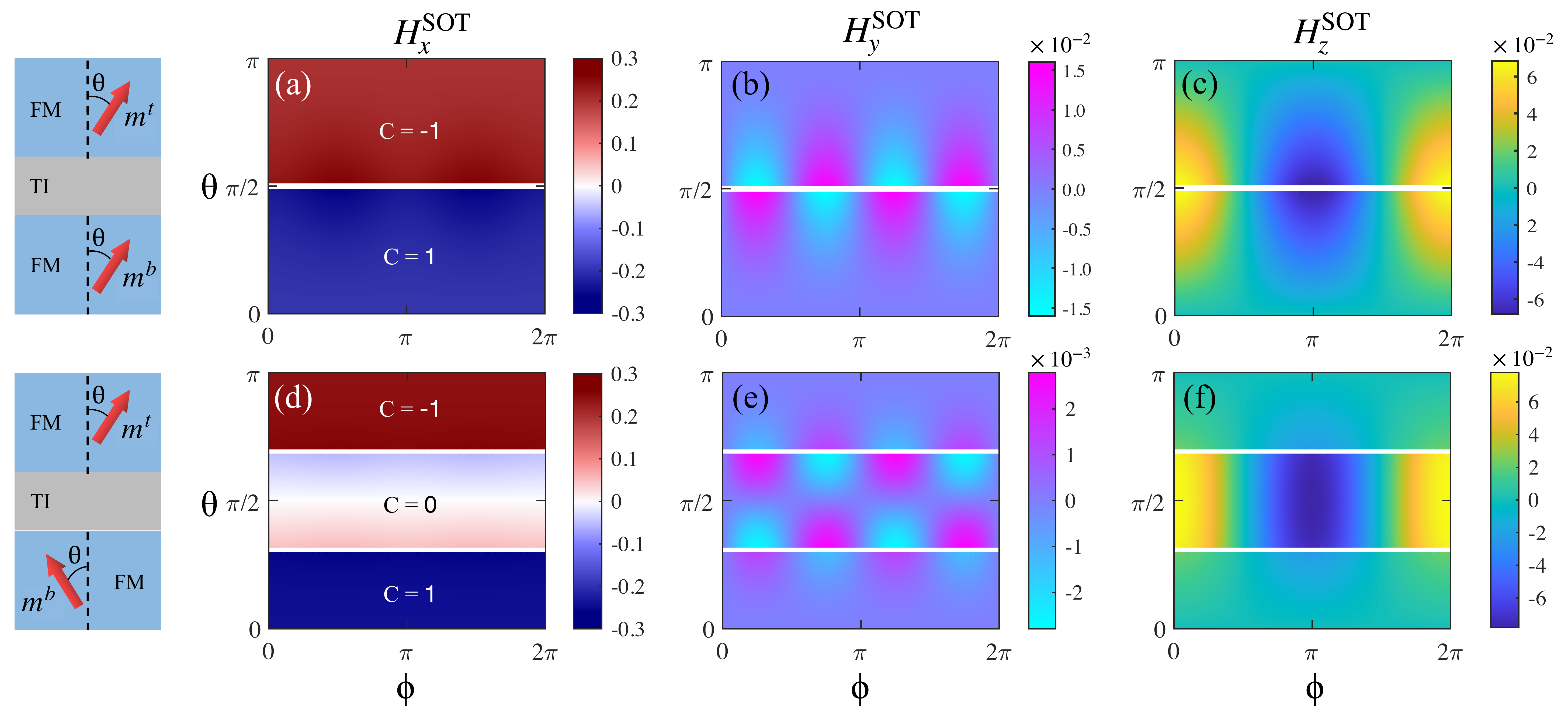}
\caption{The SOT field acting on the top FM (in $x$, $y$ and $z$ components) as a function of the polar angle $\theta$ and the azimuth angle $\phi$ when $\bm{m}^t$ and $\bm{m}^b$ are (a)-(c) in phase, and (d)-(f) in the exchange mode configuration. The coordinate systems are illustrated in the left panels. The blank strips mark the boundary dividing distinct topological phases, on which the band gaps close. $\bm{H}^{\rm SOT}_t$ is plotted in the unit $(ea_0 E_x)/(4\pi^2N_z\hbar S)$ (frequency dimension) where $E_x$ is the applied electric field. Parameters: $J_{sd}=100 \rm meV$, $b_{0}=50 \rm meV$, $b_{1}=1 \rm eV\cdot nm^2$ and $\hbar v_f=1 \rm eV\cdot nm$.}
\label{fig:SOT} 
\end{figure*}

We now apply the formalism above to a FM-TI-FM trilayer system as schematically illustrated in Fig.~\ref{fig:model}(a), assuming that both FM layers are insulating and having the same thickness (so $N^t_z=N^b_z=N_z$). On the interfaces, the topological electrons couple to the adjacent magnetic moments via the exchange interaction $J_{sd}$. Under the basis $(c^{t}_{\bm{k}\uparrow},c^{t}_{\bm{k}\downarrow},c^{b}_{\bm{k}\uparrow},c^{b}_{\bm{k}\downarrow})^T$ where $c^{t(b)}_{\bm{k}\sigma}$ annihilates an electron with wave vector $\bm{k}=\{k_x,k_y\}$ and spin $\sigma$ at the top (bottom) interface, the Hamiltonian $H_0$ in Eq.~\eqref{eq:general Hamiltonian} can be written as~\cite{Y.Rui_Science_2010,Ali_2020_PRL}:
\begin{align}
\label{eq:Hamiltonian}
H_0=& \hbar v_{f}(k_x \tau_z\otimes\sigma_y -k_y \tau_z\otimes \sigma_x)+A(\bm{k})\tau_x\otimes I \notag\\
&\quad +J_{sd}(\bm{m}^t \cdot \bm{\sigma})\oplus (\bm{m}^b \cdot \bm{\sigma}),
\end{align}
where $\otimes$ ($\oplus$) denotes direct product (direct sum), $v_f$ is the Fermi velocity, $\bm{\sigma}$ and $\bm{\tau}$ are the vectors of Pauli matrices representing the spin and layer (\textit{i.e.}, top and bottom) degrees of freedom, respectively~\cite{YuHangLi_PRL_2021}. $A(\bm{k})=b_0+b_1(k_{x}^2+k_{y}^2)$ describes the overlap between the top and bottom surface states in the thin TI film, where  $b_0$ and $b_1$ depend on the TI thickness.

We first look into a simple case where the two FM layers are collinear (but not necessarily along the $\hat{z}$ direction) and are described by a single magnetization vector $\bm{m}$. The energy dispersion can then be solved from Eq.~\eqref{eq:Hamiltonian} as
\begin{align}
\varepsilon = \pm\sqrt{A(\bm{k})^2+(\hbar v_f \bm{k})^2+J^2_{sd}\pm\xi}\label{eq:band},
\end{align}
where
\begin{align}
 \xi=2J_{sd}\sqrt{(\hbar v_f)^2[\bm{\hat{z}}\cdot(\bm{k}\times\bm{m})]^2+A(\bm{k})^2}.
\end{align}
Figure~\ref{fig:model}(b) shows the band structure around the $\Gamma$ point, where aside from the exchange gap determined by $J_{sd}$, the overlap between the top and bottom interfaces $b_0$ opens a hybridization gap $\Delta_2\propto b_0$. There exists a topological phase transition at $b_0=J_{sd}$~\cite{Y.Rui_Science_2010,Yunyou.Yang_APL_2011}, where the band gap $\Delta_1\propto (J_{sd}-b_0)$ closes. The topologically nontrivial (trivial) regime corresponds to $b_0<J_{sd}$ ($b_0>J_{sd}$).

To obtain the effective SOT field $\bm{H}^{\rm SOT}$ defined in Eq.~\eqref{eq:SOT} as a function of the $\bm{m}$ direction, we now perform numerical calculation using typical material parameters: $J_{sd}=100$ $\rm meV$, $b_{0}=50$ $\rm meV$, $b_{1}=1$ $\rm eV nm^2$ and $\hbar v_f=1$ $\rm eV nm$~\cite{H.J.Zhang_NatPhys_2009,Lu_PRB_2010,Linder_PRB_2009}. The Fermi level lies in the surface gap in the absence of doping and gating. Figures~\ref{fig:SOT}(a)-(c) plot the angular dependence of $\bm{H}^{\rm SOT}$ exerting on $\bm{m}^t$, while $\bm{H}^{\rm SOT}$ acting on $\bm{m}^b$ is exactly the opposite (thus not plotted). The corresponding SOT is $\tau=\gamma\bm{H}^{\rm SOT}\times\bm{m}$. We observe that $\bm{H}^{\rm SOT}$ is largely pointing in the $\hat{x}$ direction, \textit{i.e.}, collinear with the applied electric field $\bm{E}$. However, different from the case of FM-TI bilayer where $\bm{H}^{\rm SOT}$ is exactly along the direction of $\bm{E}$, here the SOT field also has small but finite projections on the $y$ and $z$ axes, the existence of which can be attributed to the surface mixing term $A(\bm{k})$ in the Hamiltonian.

Because $\bm{H}^{\rm SOT}$ is opposite for $\bm{m}^t$ and $\bm{m}^b$, the two FM initially polarized along $\bm{z}$ will evolve towards opposite directions, which subsequently brings about noncollinearity and triggers the interlayer exchange interaction, leading to a precessional motion of the two FM with a $\pi$-phase difference. This means that the high-frequency exchange mode can be excited by an AC electric field through the voltage-induced SOT. However, in the exchange mode, $\bm{m}^t$ and $\bm{m}^b$ are noncollinear and have opposite in-plane components. Therefore, the band structure, the Berry curvature, hence $\bm{H}^{\rm SOT}$, all depending on the spin configuration, must be re-calculated based on the new configuration: $m^t_x=-m^b_{x}$, $m^t_y=-m^b_y$ and $m^t_z=m^b_z$. Figures~\ref{fig:SOT}(d)-(f) plot $\bm{H}^{\rm SOT}$ for the exchange mode, where we find that $H^{\rm SOT}_{x}$ is still the dominant component. Just as in the collinear configuration, $\bm{m}^t$ and $\bm{m}^b$ are subject to opposite SOT fields. So, for small-angle precessions, the voltage-induced SOT is able to sustain the exchange mode, which will be further discussed in Sec.~\ref{sec:EMR}.

Comparing to Fig.~\ref{fig:SOT}(a) in which two topological phases with Chern number $C=\pm1$ are adjacent (given that $b_0<J_{sd}$), a topological trivial phase ($C=0$) is identified in Fig.~\ref{fig:SOT}(d) between $\theta=55^\circ$ and $\theta=125^\circ$ (these values depend on $b_0$). In the $C=0$ phase, $H_{x}^{\rm SOT}$ is small but not zero, suggesting that the voltage-induced SOT involves the bulk (surface) effect and is not a pure edge effect. By contrast, the $H_{y}^{\rm SOT}$ and $H_{z}^{\rm SOT}$ components are larger in the $C=0$ phase than in the $C=\pm1$ phases. While both $H_{y}^{\rm SOT}$ and $H_{z}^{\rm SOT}$ exhibit sinusoidal dependence on $\phi$, the former (latter) has a period of $\pi$ ($2\pi$). Moreover, $H_{y}^{\rm SOT}$ ($H_{z}^{\rm SOT}$) is an odd (even) function of $(\theta-\pi/2)$.

To conclude this section, we finally evaluate the "torkance" defined as the SOT field per applied electric field: $\gamma H^{\rm SOT}/E$. With the parameters used in plotting Fig.~\ref{fig:SOT}, the torkance is about $1$ $\rm GHz/(V/\mu m)$ for $N_zS=10$ (see Appendix~\ref{sec:appendix dissipative coupling}). As a comparison, in a typical current-driven ferromagnetic resonance~\cite{L.Q.Liu_PRL_2017,STFMR_Liu}, the torkance is on the same order of magnitude when platinum (Pt) is used as the spin generator.

\section{SOT-Driven Exchange Mode Resonance}\label{sec:EMR}

\begin{figure}[t]
\centering
\includegraphics[width=0.95\linewidth]{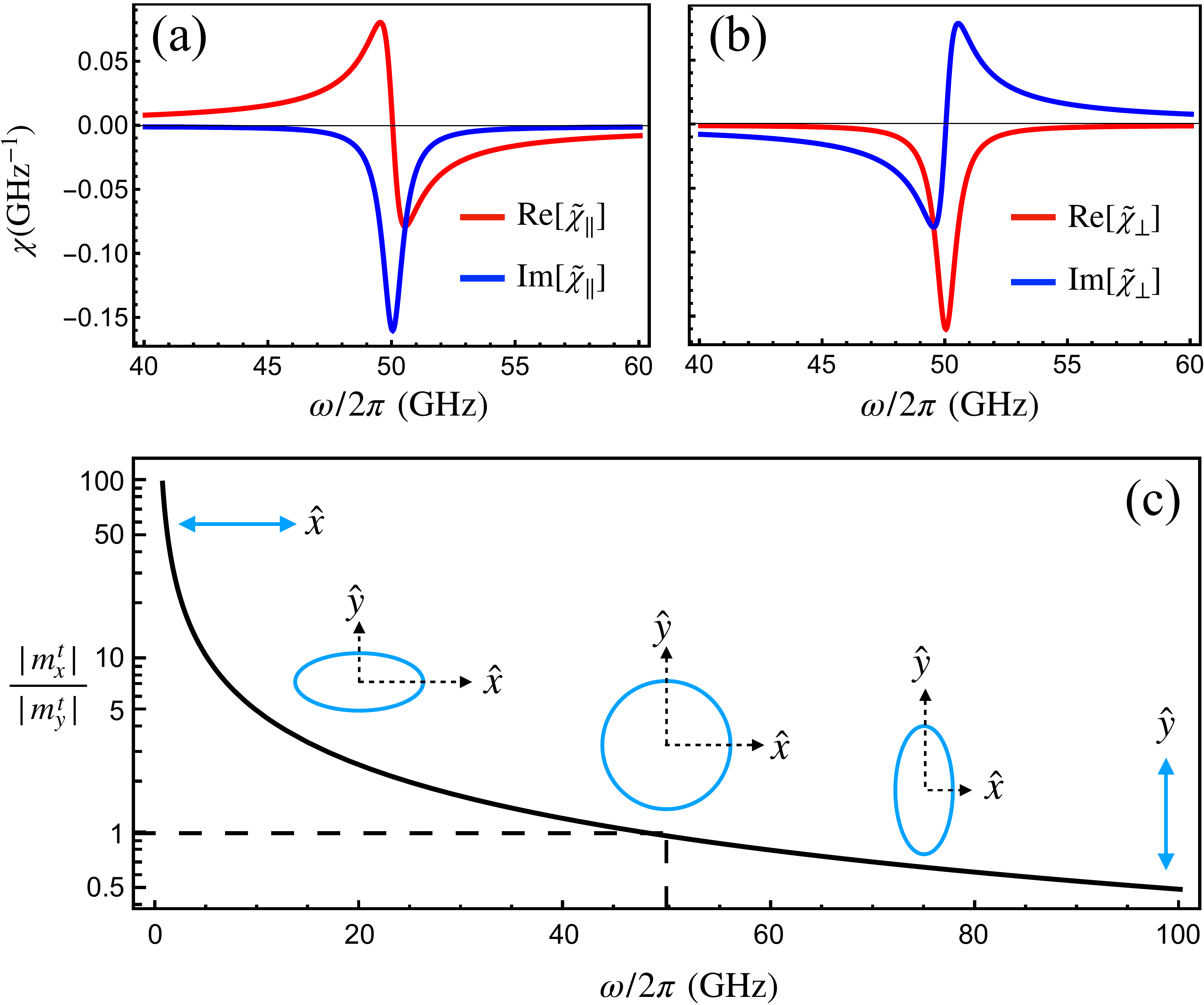}
\caption{Real and imaginary parts of the dynamical susceptibility for: (a) parallel component $\tilde{\chi}_\parallel$ and (b) perpendicular component $\tilde{\chi}_\perp$ as functions of $\omega$ with a resonance frequency $\omega_r/2\pi=50\ \rm GHz$ and a driving SOT field $\gamma H^{\rm SOT}_{x}=0.628$ GHz. (c) Logarithmic plot of $|m^t_{x}/m^t_{y}|$, where at the EMR point (dashed line) the ratio is $1$. Schematics of the polarization is illustrated at different frequencies.}
\label{fig:EMR}
\end{figure}

To study the excitation of EMR by the voltage-induced SOT derived above, we resort to the coupled Landau–Lifshitz–Gilbert (LLG) equations:
\begin{subequations}
\label{eq:LLG}
\begin{align}
\dot{\bm{m}}^t&=(\omega_E \bm{m}^b+\omega_A m^t_z\hat{\bm{z}}+\bm{H}^{\rm SOT})\times\bm{m}^t+\alpha\bm{m}^t\times\dot{\bm{m}}^t\label{eq:EMR_LLGa},\\
\dot{\bm{m}}^b&=(\omega_E \bm{m}^t+\omega_A m^b_z\hat{\bm{z}}-\bm{H}^{\rm SOT})\times\bm{m}^b+\alpha\bm{m}^b\times\dot{\bm{m}}^b\label{eq:EMR_LLGb},
\end{align}
\end{subequations}
where $\omega_A$ and $\omega_E$ are the angular frequencies of the easy-axis anisotropy and interlayer exchange interaction, respectively, and $\alpha$ is the dimensionless Gilbert damping constant. Here we assume the two FM layers are identical and share the same anisotropy, but our theory can easily be generalized to account for non-equivalent FM layers. While $\omega_E$ may be either ferromagnetic or antiferromagnetic depending on the TI thickness~\cite{MingDa.Li_PRB_2015,B.A.Mansoor_AIPAdv_2017}, here we assume that $\omega_E>0$ (\textit{i.e.}, ferromagnetic coupling). As we are considering insulating FM and gapped surface states, there is no transport current and the Oersted field in ordinary spin-torque resonance experiments does not exist. Nonetheless, the applied AC electric field amounts to a displacement current proportional to $\partial\bm{E}/\partial t$, which, according to Maxwell's equation, generates a stray field similar to the Oersted field. We estimate that the stray field stemming from the displacement current is negligible for thin films (see Appendix~\ref{sec:appendix dissipative coupling}), thus to a good approximation, only the SOT field $\bm{H}^{\rm SOT}$ needs to be taken into consideration. Linearizing the LLG equations around the ground state $\bm{m}^t=\bm{m}^b=\hat{\bm{z}}$ with respect to the phase vectors $\tilde{m}_{\perp}^{t(b)}=(\tilde{m}^{t(b)}_x, \tilde{m}^{t(b)}_y)e^{i\omega t}$, we obtain $\tilde{m}^{t/b}_x=\pm\tilde{\chi}_{\parallel} \tilde{H}_x^{\rm SOT}$ and $\tilde{m}^{t/b}_y=\pm\tilde{\chi}_{\perp}\tilde{H}_x^{\rm SOT}$. Here, the driving field $\tilde{H}_x^{\rm SOT}=H_x^{\rm SOT}e^{i\omega t}$ while the small $H_{y}^{\rm SOT}$ and $H_{z}^{\rm SOT}$ components are ignored, and
\begin{subequations}
\begin{align}
\tilde{\chi}_{\parallel}(\omega)=\frac{- (i\alpha\omega+\omega_r)}{\omega^2-(i\alpha\omega+\omega_r)^2} \label{eq:chix},\\
\tilde{\chi}_{\perp}(\omega)=\frac{i\omega }{\omega^2-(i\alpha\omega+\omega_r)^2} \label{eq:chiy},
\end{align}
\end{subequations}
where $\omega$ is the driving frequency of the applied electric field, $\omega_{r}=2\omega_E+\omega_A$ is the bare EMR frequency for negligible $\alpha$~\cite{Unequallayers}. Depending on $\omega_{E}$, the EMR frequency could be much higher than that of ordinary ferromagnetic resonances which is on the order of $\omega_A$ in the absence of external magnetic fields. Figures~\ref{fig:EMR}(a) and (b) plot $\tilde{\chi}_{\parallel}$ and $\tilde{\chi}_{\perp}$ for an EMR frequency of $\omega_r/2\pi=50$ GHz, where we see that the real (imaginary) part of $\tilde{\chi}_{\parallel}$ is antisymmetric (symmetric) around $\omega_r$ and the symmetry pattern of $\tilde{\chi}_{\perp}$ is just the opposite. For very small $\alpha$, the phase difference between $\tilde{\chi}_{\parallel}$ and $\tilde{\chi}_{\perp}$ is about $\pi/2$ and almost independent of frequency (unless $\omega/\omega_r$ is as large as $1/\alpha$), which indicates a persistent counterclockwise precession of $\bm{m}^{t(b)}$ regardless of the driving frequency, as illustrated in Fig.~\ref{fig:EMR}(c). Nonetheless, when going off-resonance, the magnitude $|\tilde{\chi}_{\parallel}|$ decays faster (slower) than $|\tilde{\chi}_{\perp}|$ for $\omega>\omega_r$ (for $\omega<\omega_r$). Therefore, the actual polarization becomes elliptical and evolves towards opposite linear oscillations when moving away from the EMR point on which a circular polarization is achieved.

\section{Topological charge pumping}\label{sec:pumping}

\begin{figure}[t]
\centering
\includegraphics[width=\linewidth]{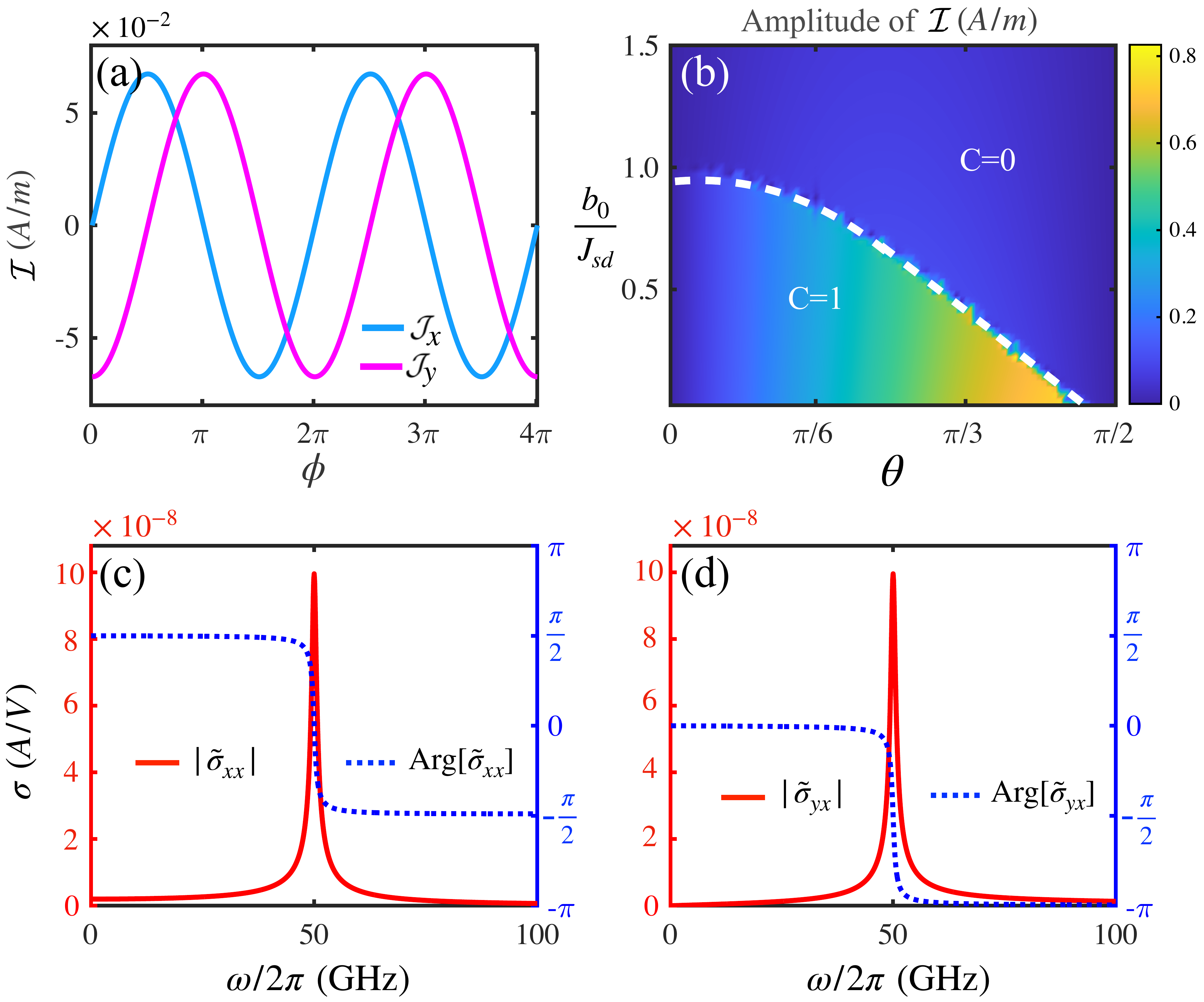}
\caption{(a) Surface current density in the $x$ and $y$ directions as a function of the precessional angle $\phi$ at the EMR for $\omega_r/2\pi=50$ GHz. For visual clarity, the amplitude of oscillation is exaggerated to $|m^t_{x(y)}|=0.1$. (b) The amplitude of surface current density $\mathcal{I}$ as a function of the polar angle $\theta$ and the ratio $b_0/J_{sd}$ ($J_{sd}=100 $ $\rm meV$). The phase boundary is marked by a white dashed line. (c) and (d): Amplitude and phase of electrical admittance $\tilde{\sigma}_{xx}$ and $\tilde{\sigma}_{yx}$ as a function of the driving frequency $\omega$. Parameters: $\alpha=0.01$ and $N_zS=10$.}
\label{fig:sigma}
\end{figure}

With the onset of EMR, the precessing magnetization generates an adiabatic current through topological charge pumping, which is the reciprocal (reverse) effect of voltage-induced SOT~\cite{D.J.Thouless_PRB_1983}. This picture remains valid so long as the adiabatic condition $\hbar\omega\ll\Delta_1$ is satisfied (\textit{i.e.}, the motion of $\bm{m}^t(t)$ and $\bm{m}^b(t)$ does not agitate inter-band transitions of electrons). In contrast to transport currents accompanied by Joule heating, an adiabatic current is non-dissipative and incurs no Ohmic loss~\cite{Oppen_PRL_2013,arrachea2015nanomagnet,D.J.Thouless_PRB_1983}, which is a salient advantage of voltage-driven systems. At the same time, topological charge pumping produces an output AC current that directly signals the magnetization dynamics, through which the EMR can be experimentally detected. Inserting $\dot{\bm{r}}_c$ expressed in Eq.~\eqref{eq:EOMb} into the surface current density $\mathcal{I}=-e/(2\pi)^2\int \dot{\bm{r}}_cf(\bm{k}_c)d\bm{k}_c$ gives~\cite{Hallcurrent}
\begin{align}
\mathcal{I}=\frac{-e}{(2\pi)^2}\int_{BZ}\left(\dot{\bm{m}}^t\cdot\overleftrightarrow{\Omega}_{m^t k}+\dot{\bm{m}}^b\cdot\overleftrightarrow{\Omega}_{m^bk}\right)d\bm{k}_c. \label{eq:pumping}
\end{align}
When being applied to a simple FM-TI bilayer, Eq.~\eqref{eq:pumping} has only one term, yields exactly the same result derived in Ref.~\cite{Y.Takehito_PRB_2012}. In the ferromagnetic mode of our FM-TI-FM trilayer, however, the in-phase precession of $\bm{m}^t$ and $\bm{m}^b$ cancel each other so that $\mathcal{I}$ vanishes. By contrast, in the exchange mode they have a $\pi$ phase difference, thus constructively contributing to the output current. At the EMR point, the system becomes circularly polarized, which reflects also in the pumped current as plotted in Fig.~\ref{fig:sigma}(a), where the $x$ and $y$ components are both sinusoidal and differ by a phase of $\pi/2$.

As the reverse effect of SOT, topological charge pumping should also depend on the amplitude of EMR characterized by $\theta$ and the band topology of electrons characterized by the Chern number. In this regard, we numerically compute $\mathcal{I}$ as a function of $\theta$ and $b_0/J_{sd}$ and plot the result in Fig.~\ref{fig:sigma}(b). We see that the pumped current vanishes in the topological trivial phase ($C=0$); the maximum ratio of $b_0/J_{sd}$ to keep the system topologically non-trivial decreases with increasing $\theta$ (\textit{i.e.}, growing $|m_\perp|$) until it vanishes at a critical value $\theta_c$. From the shape of the phase boundary in Fig.~\ref{fig:sigma}(b), we read off a good tolerance on the resonance amplitude: the system remains in the $C=1$ phase even when $\theta$ is no longer small at EMR unless $b_0/J_{sd}\rightarrow1$. For example, if $b_0=0.8J_{sd}$, $\theta$ can grow up to $\pi/6$ (which is certainly not a small-angle) without losing the band topology.

Because the SOT and the topological charge pumping take place simultaneously, their combined effect manifests as a linear relation between the output current density and the driving electric field
\begin{align}
 \tilde{\mathcal{I}}_\mu(\omega)=\tilde{\sigma}_{\mu\nu}(\omega)\tilde{E}_\nu(\omega),
\end{align}
where $\tilde{\sigma}_{\mu\nu}$ is the electrical admittance (or AC conductivity). Substituting the LLG equations~\eqref{eq:LLG} into Eq.~\eqref{eq:pumping} to eliminate $\dot{\bm{m}}^b$ and $\dot{\bm{m}}^t$, and considering Eq.~\eqref{eq:SOT}, we obtain:
\begin{align}
    \tilde{\sigma}_{xx}(\omega)=i\zeta\omega\tilde{\chi}_\parallel(\omega)\overline{\Omega}_{k_x m^t_x}^2, \label{eq:sigmaxx}\\
    \tilde{\sigma}_{yx}(\omega)=i\zeta\omega\tilde{\chi}_\perp(\omega)\overline{\Omega}_{k_y m^t_y}^2, \label{eq:sigmayx}
\end{align}
where $\zeta=\frac{2e^2}{N_z\hbar Sa_0^2}$ (see details in Appendix~\ref{sec:appendix dissipation}). The amplitude and phase of $\tilde{\sigma}_{xx}(\omega)$ and $\tilde{\sigma}_{yx}(\omega)$ are plotted in Figs.~\ref{fig:sigma}(c) and (d), where the amplitude of both components exhibits a sharp peak around the EMR point $\omega_r=50\ {\rm GHz}$. Since $\zeta$ scales inversely with $N_zS$, the thicker the FM films, the smaller the peak value of electrical admittance. As a comparison, we note that the quantized Hall conductivity $\sigma_H=e^2/h$ is on the order of $10^{-5} {\rm A/V}$, so the peak value of $\tilde{\sigma}_{yx}(\omega)$ for $N_zS=10$ is about two orders of magnitude smaller. Therefore, the transverse component of the pumped current is overwhelmed by the anomalous Hall effect, and is thus very difficult to detect.

\begin{figure}[t]
\includegraphics[width=0.8\linewidth]{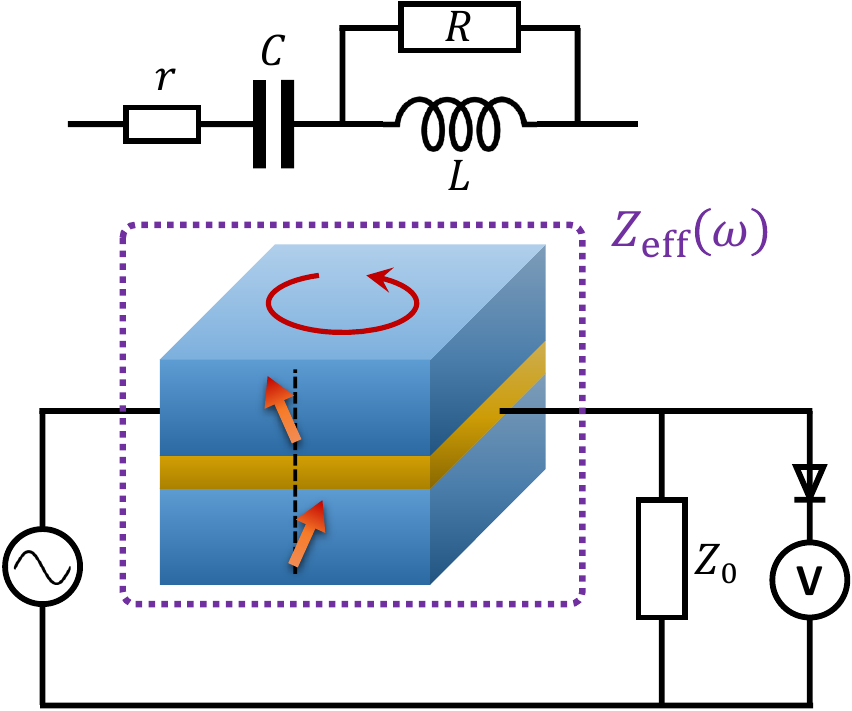}
\caption{Detection scheme of the EMR and effective circuit for the FM-TI-FM trilayer. $Z_0$ is a fixed impedance, the voltage drop on which is monitored by a voltmeter.}
\label{fig:device} 
\end{figure}

By contrast, the longitudinal output current determined by $\tilde{\sigma}_{xx}(\omega)$ is a conspicuous effect, because the whole system is insulating and the output current solely originates from the magnetization precessions. Basing on the form of $\tilde{\sigma}_{xx}(\omega)$, we conceive an experimental scheme as illustrated in Fig.~\ref{fig:device} to measure topological charge pumping, and hence the EMR, of the FM-TI-FM trilayer heterostructure. We connect the trilayer device in series with a reference impedance $Z_0$ and power the whole circuit by an AC voltage source. The voltage drop across $Z_0$ measured by a voltmeter vanishes (reaches maximum) for off-resonance (on-resonance) conditions as the impedance of our device, $1/\tilde{\sigma}_{xx}(\omega)$, goes to infinity (reaches minimum). To better characterize the electrical response associated with the voltage-driven EMR, we further map our device to an effective circuit whose longitudinal admittance is identical to $Z_{\rm eff}(\omega)=1/\tilde{\sigma}_{xx}(\omega)$. The effective circuit, as illustrated in the top panel of Fig.~\ref{fig:device}, consists of a capacitor, an inductor, and two resistors. The physical parameters in the effective circuit are solved as (see Appendix~\ref{sec:appendix circuit})
\begin{align}
   R=\frac{Z_m}{\alpha},\quad  r=Z_m\alpha,\quad L=\frac{Z_m}{\omega_r},\quad C=\frac{1}{Z_m\omega_r}, \label{eq:RrLC}
\end{align}
where
\begin{align}
 Z_m^{-1}=\overline{\Omega}_{k_x m^t_x}^2\zeta \frac{w}{l}
\end{align}
with $l$ and $w$ the length and width of the FM-TI-FM trilayer. For a square sample with $w/l=1$, $Z_m$ is about $5\times 10^8 {\rm \Omega}$. For Gilbert damping $\alpha=0.01$ and resonance frequency $\omega_r/2\pi=50\ {\rm GHz}$, the two resistances, the inductance, and the capacitance appearing in the effective circuit are evaluated to be $R=5\times 10^{10}\ {\rm \Omega}$, $r=5\times 10^6\ {\rm \Omega}$, $L=1.59\times 10^{-3}\ {\rm H}$ and $C=0.64\times 10^{-8}\ {\rm pF}$, respectively. It is easy to see from the effective circuit that our trilayer device behaves as a capacitor (an inductor) for $\omega\ll\omega_r$ ($\omega\gg\omega_r$); right at the EMR point, the device behaves as a resistor because the impedance from $C$ and $L$ cancel. We emphasize that $R$ and $r$ in Eq.~\eqref{eq:RrLC} do not produce real Joule heating; they simply represent magnetic dissipation (Gilbert damping) in the form of effective resistors.

\section{Mechanical Efficiency}\label{sec:mecheff}

An adiabatic quantum motor has a theoretical mechanical efficiency of $100\%$ because its output current is purely adiabatic, incurring zero Ohmic loss (Joule heating)~\cite{Oppen_PRL_2013,arrachea2015nanomagnet}. To confirm that our considered FM-TI-FM trilayer system is indeed operated as an adiabatic quantum motor, we need to quantify how much input power is eventually converted to the magnetization dynamics, which is reflected by the mechanical efficiency $\eta\equiv P_M/P_J$, where $P_M=2\alpha\overline{|\dot{\bm{m}}^2|}M_s V_m/\gamma$ is the magnetic dissipation rate with $V_m$ the volume of each FM layer ($V_m=a_0^3N_xN_yN_z$) and $P_J=\overline{|\mathcal{I}_x E_x|}wl$ is the total electric power consumed by the trilayer device. The overline here represents a time average over one period of oscillation (\textit{e.g.}, $\overline{\mathcal{I}_xE_x}\equiv\int_0^T\mathcal{I}_xE_x dt/T$ with $T=2\pi/\omega$). In the absence of leakage currents, we obtain after a lengthy derivation (see details in Appendix~\ref{sec:appendix dissipation})
\begin{align}
    P_M&=\alpha\frac{E_x^2wl\zeta}{2} \overline{\Omega}_{k_x m^t_x}^2 \omega^2\left[|\tilde{\chi}_\parallel(\omega)|^2+|\tilde{\chi}_\perp(\omega)|^2\right], \label{eq:avepm}\\
    P_J&=\frac{E_x^2wl\zeta}{2} \overline{\Omega}_{k_x m^t_x}^2 \omega \left|{\rm Im}\left[\tilde{\chi}_\parallel(\omega)\right]\right|, \label{eq:avepJ}
\end{align}
which, when Eqs.~\eqref{eq:chix} and~\eqref{eq:chiy} are inserted, leads to
\begin{align}
   \eta = \frac{P_M}{P_J} = 1, \label{eq:efficiency}
\end{align}
confirming that the considered setup indeed functions as an adiabatic quantum motor. Here, the mechanical efficiency should not be confused with the “spin-torque efficiency” defined exclusively for current-induced torques\cite{PaiChiFeng_APL_2021}.

While the ideal $100\%$ mechanical efficiency is an intrinsic characteristic of adiabatic quantum motor, real systems always have imperfections that inevitably jeopardize the mechanical efficiency. In our case, the TI may not be a perfect insulator even when the surface states are gapped by the adjacent magnetization~\cite{kim2012surface}, resulting in leakage currents that do incur Ohmic loss. The leakage effect can be represented by a very large resistance $R_{\rm leak}$ in parallel with a very small capacitance $C_{\rm leak}$, which should be added in parallel to the effective circuit drawn in Fig.~\ref{fig:device}. Under an AC voltage drive, however, $C_{\rm leak}$ does not contribute to any time-averaged dissipation. Therefore, the leakage-induced Ohmic loss makes up a dissipation power $P_{\rm leak}=E_x^2l^2/2R_{\rm leak}$, hence the actual mechanical efficiency becomes $\eta=P_M/(P_J+P_{\rm leak})$. Using typical material parameters (\textit{e.g.}, Bi$_2$Te$_2$Se~\cite{Ren_PRB_2010} for the TI and CrI$_3$ for the FM~\cite{hou2019magnetizing}), we estimate $P_{\rm leak}$ to be somewhere between $10\%$ to $50\%$ of $P_J$ depending on several geometrical factors of the device (such as the TI thickness), so the actual mechanical efficiency is still very high. With improved material qualities (\textit{e.g.}, more insulating TI) and device engineering, we are optimistic that $\eta$ can be pushed towards its theoretical limit (\textit{i.e.} $100\%$) in the future.

To benchmark the radical boost of mechanical efficiency by using voltage-induced SOT, we now estimate $\eta$ for the well-established current-driven resonances. Consider a YIG/Pt bilayer~\cite{Zhou_Xiao_2013PRB,Chiba_PRApplied_2014} where the SOT is generated by the spin Hall effect in the Pt. As calculated in Appendix~\ref{sec:appendix dissipation}, $\eta$ is below $1\%$ at the ferromagnetic resonance point~\cite{note:eta}. In current-driven systems, the mechanical efficiency is limited by three major factors: (1) only a small fraction of the driving current is converted into spin current by the spin Hall effect; (2) the backflow effect~\cite{Spin_Battery,Spin_Backflow} suppresses the effective spin-mixing conductance governing the spin transmission across interfaces; (3) the Ohmic loss in a metallic spin generator such as Pt is enormous. If the FM layer is changed into a permalloy, the shunting current effect will further reduce $\eta$ down to the order of $0.1\%$. By sharp contrast, none of these mechanisms exist in our system. First, the voltage-induced SOT does not involve bulk spin diffusion as in the spin Hall effect. Second, spins do not transmit across the interface but drive the magnetization directly as an adiabatic quantum motor. Third, the output current stems from the precessing motion of magnetization (\textit{i.e.}, topological charge pumping) which does not generate Joule heating as that for a dissipative current.

\section{Concluding Remarks}\label{sec:Summary}

In conclusion, an insulating FM-TI-FM trilayer heterostructure can be operated as an adiabatic quantum motor via the combined effect of voltage-induced spin-orbit torque and its reverse effect, topological charge pumping, which offers a theoretical mechanical efficiency of $100\%$ as Joule heating is obviated. This mechanism is particularly suitable for achieving the high-frequency exchange resonance in which the two FM layers are out of phase by $\pi$. Even in the presence of leakage currents, our proposed setup can still function with a mechanical efficiency two orders of magnitude higher than that in current-driven resonances. We anticipate that our findings will facilitate the development of ultrafast spintronic devices with extremely low-energy dissipation.

If the two FM layers are antiferromagnetically aligned, the instantaneous SOT acting on $\bm{m}^t$ is opposite to that acting on $\bm{m}^b$, whereas the associated SOT field $\bm{H}^{\rm SOT}$ is of the same sign. Consequently, an AC voltage drive can excite the antiferromagnetic resonance in such a system. Since the antiferromagnetic resonance is composed of two degenerate chiral modes in the absence of magnetic fields, the topological charge pumping, hence the overall electrical response embedded in $\tilde{\sigma}_{\mu\nu}$, becomes more complicated than that in the EMR. This will be left for future studies.

While we have demonstrated our findings in an FM-TI-FM trilayer, the essential mechanism is equally applicable to an insulating FM-TI bilayer, where the resonance frequency is lower than the EMR studied above.

We also mention that our calculations can be generalized to antiferromagnet-TI interfaces so long as the interface is uncompensated (\textit{e.g.}, using the layered antiferromagnet CrI$_3$). However, our model is invalid for compensated interfaces (\textit{e.g.}, using MnPS$_3$) where the compensating magnetic order cannot open gaps of the Dirac electrons.

The authors acknowledge fruitful discussions with H. Zhang, Y. Li and E. der Barco. This work is supported by the Air Force Office of Scientific Research under Grant No. FA9550-19-1-0307.

\appendix
\renewcommand\thefigure{B\arabic{figure}}
\setcounter{figure}{0}

\section{Lagrangian and equation of motion}\label{sec:appendix EOM}
We first verify that $\mathcal{L}_j$ gives rise to the correct LLG for $\bm{m}^j$. The magnetization vector can be parametrized by the polar angle $\theta$ and azimuthal angle $\phi$ as $\bm{m}^j=(\sin\theta^j\cos\phi^j,\sin\theta^j\sin\phi^j,\cos\theta^j)$. In spherical coordinates,
\begin{align}
    \frac{\delta \mathcal{L}}{\delta \theta}&=\cos\theta\left(\cos\phi\frac{\delta \mathcal{L}}{\delta m_x}+\sin\phi\frac{\delta \mathcal{L}}{\delta m_y}\right)-\sin\theta\frac{\delta \mathcal{L}}{\delta m_z}=\hat{e}_\theta \cdot \frac{\delta \mathcal{L}}{\delta \bm{m}},\\
    \frac{\delta \mathcal{L}}{\delta \phi}&=\sin\theta\left(-\sin\phi\frac{\delta \mathcal{L}}{\delta m_x}+\cos\phi\frac{\delta \mathcal{L}}{\delta m_y}\right)=\sin\theta \hat{e}_\phi\cdot \frac{\delta \mathcal{L}}{\delta \bm{m}},
\end{align}
where the unit vectors $\hat{e}_\theta=(\cos\theta\cos\phi,\cos\theta\sin\phi,-\sin\theta)$ and $\hat{e}_\phi=(-\sin\phi,\cos\phi,0)$. Consequently,
\begin{align}
    0=&\frac{\delta \mathcal{L}_j}{\delta \bm{m}^j}= \frac{\delta \mathcal{L}_j}{\delta \theta^j}\hat{e}_\theta+\frac{1}{\sin\theta^j}\frac{\delta \mathcal{L}_j}{\delta \phi^j}\hat{e}_\phi \notag\\
    =&\left(\frac{\partial \mathcal{L}_j}{\partial \theta^j}-\frac{d}{dt}\frac{\partial \mathcal{L}_j}{\partial \dot{\theta}^j}\right)\hat{e}_\theta + \left(\frac{\partial \mathcal{L}_j}{\partial \phi^j}-\frac{d}{dt}\frac{\partial \mathcal{L}_j}{\partial \dot{\phi}^j}\right)\frac{\hat{e}_\phi}{\sin\theta^j}. \label{eq:ELeq}
\end{align}
Applying Eq.~\eqref{eq:ELeq} to $\mathcal{L}_j=N_j(-\hbar S\dot{\phi}^j\cos\theta^j-\epsilon_j)$ gives
\begin{align}
    -\hbar S\bm{m}^j\times\dot{\bm{m}}^j-\frac{\partial \epsilon_j}{\partial \bm{m}^j}=0,\label{seq:LLG}
\end{align}
where $\dot{\bm{m}}^j=\dot{\theta}^j\hat{e}_\theta+\dot{\phi}^j\sin\theta^j\hat{e}_\phi$. By multiplying $\bm{m}^j\times$ on both sides, Eq.~\eqref{seq:LLG} reproduces the LLG equation
\begin{align}
\dot{\bm{m}}^j=\gamma\frac{\partial\varepsilon_j}{\hbar S\partial\bm{m}^j}\times \bm{m}^j.
\end{align}
Note that in our convention, $\gamma$ is positive so the magnetic moment is $-\gamma\bm{S}^j$ with $\bm{S}^j=-\hbar S \bm{m}^j$ ($S$ is the quantum number of total angular momentum). Then, by applying the Euler-Lagrangian equations to the total Lagrangian $\mathcal{L}=\mathcal{L}_e+\sum_i \mathcal{L}_i$ with the $\mathcal{L}_i$ part taken care of by Eq.~\eqref{eq:ELeq} and following the standard procedure in Ref.~\cite{Q.Niu_PRB_1999,B.G.Xiong_PRB_2018}, we obtain Eq.\eqref{eq:eom}.

For uniform motion of a  three-dimensional FM driven by  two-dimensional electrons, a factor $N^j_z$ appears in the SOT term in Eq.~\eqref{eq:effectiveLLG} because the transferred spin angular momenta is evenly distributed to all magnetic moments. The renormalized gyromagnetic ratio Eq.~\eqref{eq:gamma_renorm} originates from the $i=j$ component from the second term on the right-hand side of Eq.~\eqref{eq:EOMc}.

\section{Dissipative coupling and displacement current}\label{sec:appendix dissipative coupling}

\begin{figure}[t]
\includegraphics[width=0.8\linewidth]{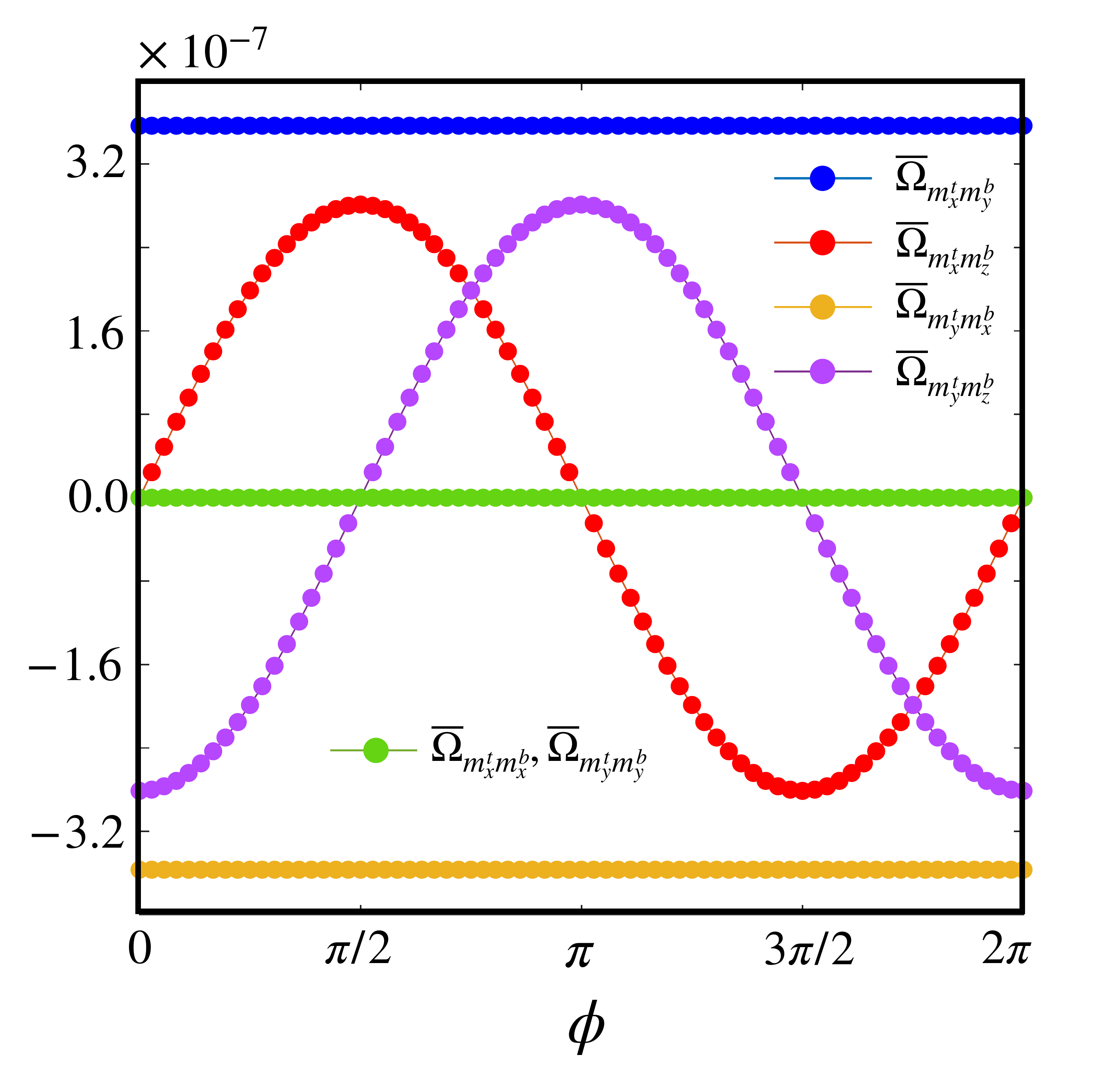}
\caption{Relevant components of $\overline{\bm{\Omega}}_{m^t m^b}$ as functions of $\phi$ for $m^t_x=-m^b_x=\sin\theta\cos\phi$, $m^t_y=-m^b_y=\sin\theta\sin\phi$ and $m^t_z=m^b_z=\cos\theta$. Parameters: $\theta=6^\circ$ and $N_z^j S=10$.}
\label{figs:dissipative} 
\end{figure}

We calculate the Berry curvature in the configurational phase space, $\overline{\bm{\Omega}}_{m^t m^b}$, with respect to angle $\phi$ for the exchange mode. The results are shown in Fig.~\ref{figs:dissipative}, where only the six relevant components are shown because of $\dot{m}_z=0$. Then the effective field stemming from $\bm{m}^t$ can be directly obtained by $\dot{\bm{m}}_t\cdot\overline{\bm{\Omega}}_{m^t m^b}$. The effective field from $\bm{m}^b$ can be obtained in a similar way. While the Gilbert $\alpha$ ranges from $10^{-2}$ to $10^{-4}$, the dissipative coupling due to Berry curvature $\overline{\Omega}_{m^t m^b}$ is on the order of $10^{-7}$ as shown in Fig.~\ref{figs:dissipative}. Therefore, we can ignore this dissipative coupling term. 

For precessions in the linear response regime, we require $|\bm{m}_{\perp}|=\sqrt{(|\tilde{m}_x|^2+|\tilde{m}_y|^2)/2}\leq0.1$. At the EMR point, we obtain from the LLG equations
\begin{align}
 |\bm{m}_{\perp}(\omega_r)|&=\gamma H^{\rm SOT}_x\sqrt{(2+\alpha^2)/[2\omega^2_r(4\alpha^2+\alpha^4)]} \notag\\
 &\approx\frac{\gamma H^{\rm SOT}_x}{2\alpha\omega_r}\leq 0.1,
\end{align}
which imposes an upper limit of the applied electric field $E_x$. For $\omega_r/2\pi=50{\rm GHz}$ and $\alpha=0.01$, the maximum value of $\gamma H^{\rm SOT}_x$ is $0.628\ {\rm GHz}$. The torkance, \textit{i.e.}, SOT field (in frequency unit) per applied electric field, is evaluated to be $e|\overline{\Omega}_{k_x m^t_x}|/N^j_z S\hbar\approx 1\ {\rm GHz/(V/\mu m)}$. Without breaking linear responses, the maximum electric field at $\gamma H^{\rm SOT}_x=0.628 {\rm GHz}$ ($|\bm{m}_\perp|=0.1$) is estimated to be $E_x=6.28\times10^5 {\rm V/m }=0.628\ {\rm V/\mu m}$.

When an AC electric field $\bm{E}$ is applied to a uniform dielectric thin film (thickness $d$ and width $w$), a displacement current is generated, which generates a magnetic field according to the Maxwell equation $\nabla \times \bm{H}=\partial_t \bm{D}=\varepsilon_0\varepsilon_r\partial_t \bm{E}$, where $\varepsilon_0$ is the vacuum permittivity and $\varepsilon_r\approx1$ is the relative permittivity. When $\bm{E}(t)=E_x(t)\hat{\bm{x}}$ and $w\gg d$ (very thin films), we have $\bm{H}(t)\approx-\hat{\bm{y}}\varepsilon_0\varepsilon_r\omega E_x(t)d/2$. Even for $d=100 \rm nm$ (thick film) and high frequency $\omega/2\pi=\omega_r/2\pi=50 {\rm GHz}$, the magnitude of the stray field is only about $|H|\approx 1.39\times10^{-2} {\rm A/m}$ when $E_x=0.628\ {\rm V/\mu m}$ (maximum value allowed, see the preceding paragraph), which is converted to a negligible magnetic flux density $|\bm{B}|=\mu_0 |\bm{H}|=1.74\times10^{-8} {\rm T}$. As a comparison, the SOT field $\gamma H^{\rm SOT}_x$ generated by the same electric field is $0.628$ GHz, which corresponds to an effective flux density of $3.569\times10^{-3}{\rm T}$ that is $5$ orders of magnitude larger than the stray field. By contrast, in current-driven resonances~\cite{STFMR_Liu}, the stray field is comparable to the SOT field.

\section{Energy dissipation and mechanical efficiency}\label{sec:appendix dissipation}

According to the LLG equation $\dot{\bm{M}}=\gamma \bm{H}^{\rm eff}\times \bm{M}+\frac{\alpha}{M_s} \bm{M}\times \dot{\bm{M}}$ with $\bm{M}=M_s \bm{m}$, the magnetic dissipation power per volume is
\begin{align}
p_m=&-\frac{d\bm{M}}{dt}\cdot \bm{H}^{\rm eff}=-\frac{\alpha}{M_s}(\bm{M}\times\dot{\bm{M}})\cdot \bm{H}^{\rm eff}\nonumber\\
=&-\frac{\alpha}{M_s} (\bm{H}^{\rm eff}\times \bm{M})\cdot\dot{\bm{M}}=-\frac{\alpha}{\gamma M_s}\dot{\bm{M}}^2. \label{seq:pm}
\end{align}
The time-averaged total magnetic dissipation power is then $P_M=2|\overline{p}_m V_m|\approx 2\alpha V_m |\overline{\dot{\bm{M}}^2}|/\gamma M_s=2\alpha M_s V_m |\overline{\dot{\bm{m}}^2}|/\gamma$ where the factor 2 accounts for both the top and bottom FMs. 

Since $E(t)={\rm Re}[E_x e^{i\omega t}]$ and $\mathcal{I}(t)={\rm Re}[\mathcal{I}_x e^{i\omega t+i\phi(\omega)}]$, the instantaneous electrical power is
\begin{align}
P_J(t)=&\mathcal{I}(t)E(t)wl=\mathcal{I}_x E_x wl\cos(\omega t)\cos(\omega t+\phi)\nonumber\\
=&\frac{\mathcal{I}_x E_x wl}{2}[\cos(2\omega t+\phi)+\cos(\phi) ].\label{seq:pJ}   
\end{align}
After taking the time average $P_J=\int_0^T P_J(t) dt/T$ with $T=2\pi/\omega$, the $\cos(2\omega t+\phi)$ term drops out, so 
\begin{align}
 P_J=\frac{\mathcal{I}_x E_x wl}2\cos(\phi)=\frac12{\rm Re}[\tilde{\mathcal{I}}_x\tilde{E}_x^*]wl,
\end{align}
where $\tilde{\mathcal{I}}_x=\mathcal{I}_xe^{i\omega t}$ and $\tilde{E}_x=E_xe^{i\omega t}$. Because $\tilde{\mathcal{I}}_x=\tilde{\sigma}_{xx}\tilde{E}_x$ and $\tilde{\sigma}_{xx}=|\tilde{\sigma}_{xx}|e^{i\phi(\omega)}$, we have
\begin{align}
P_J=\frac{E_x^2 wl}{2}|\tilde{\sigma}_{xx}|\cos(\phi)=\frac{E_x^2 wl}{2}{\rm Re}[\tilde{\sigma}_{xx}]. \label{seq:pj_cond}
\end{align}
Note that, at resonance $\tilde{\sigma}_{xx}$ is real and $\tilde{\sigma}_{xx}=10^{-7} \rm A/V$ according to Fig.~\ref{fig:sigma}(c). With typical device geometry $w=l=10\ {\rm \mu m}$ and $E_x=6.28\times 10^{5} {\rm V/m}$, $P_J$ is estimated to be at the order of $10^{-6}\rm W$ which is three order smaller than that of current-induced ST-FMR system, indicating a relatively low heating effect.

Next we relate $P_M$ and $P_J$ to the Berry curvature $\overline{\bm{\Omega}}_{km}$. Consider $\tilde{m}_{\perp}^{t}=(\tilde{m}^{t}_x,\  \tilde{m}^{t}_y)e^{i\omega t}$ where $\tilde{m}^{t}_x=\tilde{\chi}_{\parallel} \tilde{H}_x^{\rm SOT}$ and $\tilde{m}^{t}_y=\tilde{\chi}_{\perp} \tilde{H}_x^{\rm SOT}$ with $\tilde{H}_{x}^{\rm SOT}=\frac{e\tilde{E}_x}{N_z\hbar S}\overline{\Omega}_{k_x m^t_x}$, we know
\begin{align}
    m_x^t(t)=&{\rm Re}[\tilde{m}^t_x]={\rm Re}[\tilde{\chi}_\parallel \tilde{H}^{\rm SOT}_x]=\frac{e\overline{\Omega}_{k_x m^t_x}}{N_z\hbar S}{\rm Re}[\tilde{\chi}_\parallel \tilde{E}_x] \nonumber\\
    =&\frac{e\overline{\Omega}_{k_x m^t_x}}{N_z\hbar S}|\tilde{\chi}_\parallel| E_x \cos(\omega t+\phi^m_\parallel), \label{seq:mtx}\\
     m_y^t(t)=&{\rm Re}[\tilde{m}^t_y]={\rm Re}[\tilde{\chi}_\perp \tilde{H}^{\rm SOT}_x]=\frac{e\overline{\Omega}_{k_x m^t_x}}{\hbar N_z S}{\rm Re}[\tilde{\chi}_\perp \tilde{E}_x] \nonumber\\
     =&\frac{e\overline{\Omega}_{k_x m^t_x}}{N_z\hbar S}|\tilde{\chi}_\perp| E_x \cos(\omega t+\phi^m_\perp), \label{seq:mty}
\end{align}
where $\phi^m_{\parallel(\perp)}=\phi^m_{\parallel(\perp)}(\omega)$ is the phase of $\tilde{\chi}_{\parallel(\perp)}(\omega)$. Considering Eqs.~\eqref{eq:chix} and Eq.~\eqref{eq:chiy}, we know the phase difference
\begin{align}
 \phi^m_\parallel-\phi^m_\perp={\rm Arg}\left[\tilde{\chi}_{\parallel}/\tilde{\chi}_{\perp}\right]={\rm Arg}\left[-\alpha+i\omega_r/\omega\right],
\end{align}
which is about $\pi/2$ and almost independent of $\omega$ unless $\omega/\omega_r$ is comparable to $1/\alpha$.  Using Eqs.~\eqref{seq:mtx} and~\eqref{seq:mty}, we know:
\begin{align}
    |\dot{\bm{m}}^t|^2= &\dot{m}_x^{t\ 2}+\dot{m}_y^{t\ 2} \notag\\
    =&\left(\frac{eE_x\omega\overline{\Omega}_{k_x m^t_x}}{N_z\hbar S}\right)^2 \sum_{\mu=\parallel,\perp}|\tilde{\chi}_\mu|^2\sin^2(\omega t+\phi^m_\mu), \label{seq:m2}
\end{align}
which, after averaging over time, gives
\begin{align}
    \overline{|\dot{\bm{m}}^t|^2}=\left(\frac{eE_x\omega\overline{\Omega}_{k_x m^t_x}}{N_z\hbar S}\right)^2 \frac{|\tilde{\chi}_\parallel|^2+|\tilde{\chi}_\perp|^2}{2}. \label{seq:avem2}
\end{align}
The same expression applies to $\bm{m}^b$ as well. Therefore, the total magnetic dissipation power is
\begin{align}
    P_M=\left(\frac{eE_x\omega\overline{\Omega}_{k_x m^t_x}}{N_z\hbar S}\right)^2 \frac{\alpha M_s V_m (|\tilde{\chi}_\parallel|^2+|\tilde{\chi}_\perp|^2)}{\gamma}, \label{seq:avepm}
\end{align}
which proves Eq.~\eqref{eq:avepm} when we substitute $M_s=\gamma S\hbar/a_0^3$ and $V_m=a_0^3N_xN_yN_z$, and $wl=N_xN_ya_0^2$.

Integrating out $\bm{k}_c$ in Eq.~\eqref{eq:pumping} gives
\begin{align}
   \mathcal{I}_x(t)=-2e/a_0^2(\dot{\bm{m}}_x^t\cdot\overline{\Omega}_{m^t_x k_x}+\dot{\bm{m}}_y^t\cdot\overline{\Omega}_{m^t_y k_x}), \label{seq:jxt}
\end{align}
where the factor $2$ ascribes to the constructive relation between the top and bottom FM, as $\dot{\bm{m}}^t=-\dot{\bm{m}}^b$ and $\overline{\Omega}_{m^t k}=-\overline{\Omega}_{m^b k}$. Numerically, we find $|\overline{\Omega}_{k_x m^t_y}|\ll|\overline{\Omega}_{k_x m^t_x}|$ for small $\theta$ (see Fig.~\ref{fig:SOT}), so the second term on the right-hand side of Eq.~\eqref{seq:jxt} can be neglected [as we neglect $H^{\rm SOT}_y$ in Eq.~\eqref{eq:LLG} and only focus on dominant $H_x^{\rm SOT}$], and then
\begin{align}
    \mathcal{I}_x(t)=-\frac{2e^2 E_x (\overline{\Omega}_{k_x m^t_x})^2 }{N_z\hbar S a_0^2} \omega|\tilde{\chi}_\parallel| \sin(\omega t+\phi^m_\parallel), \label{seq:jxtBerry}
\end{align}
from which we can read off the conductivity
\begin{align}
    \tilde{\sigma}_{xx}=\frac{2ie^2  (\overline{\Omega}_{k_x m^t_x})^2 }{N_z\hbar S a_0^2}\omega\tilde{\chi}_\parallel. \label{seq:sigmaxx}
\end{align}
Likewise, by finding $\mathcal{I}_y(t)$ from Eq.~\eqref{seq:jxt}, we obtain
\begin{align}
 \tilde{\sigma}_{yx}=\frac{2ie^2  (\overline{\Omega}_{k_y m^t_y})^2 }{N_z\hbar S a_0^2}\omega\tilde{\chi}_\perp, \label{seq:sigmayx}
\end{align}
which proves Eqs.~\eqref{eq:sigmaxx} and~\eqref{eq:sigmayx}. Inserting Eq.~\eqref{seq:sigmaxx} into Eq.~\eqref{seq:pj_cond} gives
\begin{align}
    P_J=&-\frac{e^2  (\overline{\Omega}_{k_x m^t_x})^2 wl}{N_z\hbar S a_0^2}\omega|\tilde{\chi}_\parallel| E^2_x \sin(\phi_\parallel^m)\nonumber\\
    =&-\frac{E_x^2wl\zeta}{2} \overline{\Omega}_{k_x m^t_x}^2 \omega {\rm Im}(\tilde{\chi}_\parallel), \label{seq:pJave}
\end{align}
which proves Eq.~\eqref{eq:avepJ} as ${\rm Im}(\tilde{\chi}_\parallel)<0$. Note that we have considered an open circuit condition in the $y$ direction, so $P_J$ does not have any contribution from $\mathcal{I}_y$.

With the expressions for $P_M$ and $P_J$, we obtain the mechanical efficiency as
\begin{align}
     \eta=\frac{P_M}{P_J}=\frac{\alpha\omega (|\tilde{\chi}_\parallel|^2+|\tilde{\chi}_\perp|^2)}{|{\rm Im}(\tilde{\chi}_\parallel)|}. \label{seq:efficiency}
\end{align}
From Eqs.~\eqref{eq:chix} and~\eqref{eq:chiy}, we know
\begin{align}
|\tilde{\chi}_{\parallel}|^2=\frac{\alpha^2\omega^2+\omega_r^2}{(1+\alpha^2)^2\omega^4+2(\alpha^2-1)\omega^2\omega_r^2+\omega_r^4}\label{seq:ampchix},\\
|\tilde{\chi}_{\perp}|^2=\frac{\omega^2}{(1+\alpha^2)^2\omega^4+2(\alpha^2-1)\omega^2\omega_r^2+\omega_r^4},\label{seq:ampchiy}\\
{\rm Im}[\tilde{\chi}_\parallel]=\frac{-\alpha\omega[(1+\alpha^2)\omega^2+\omega_r^2]}{(1+\alpha^2)^2\omega^4+2(\alpha^2-1)\omega^2\omega_r^2+\omega_r^4}. \label{seq:imchix}
\end{align}
inserting which into Eq.~\eqref{seq:efficiency} yields $\eta=1$.

As a comparison, in current-driven magnetization dynamics, $P_M$ and $P_J$ are both proportional to the driving current density $j_c$ which can be fixed by external circuits. For example, in a YIG/Pt bilayer, the spin Hall effect converts $j_c$ into a pure spin current density $j_s$ flowing in the perpendicular direction. Regardless of the magnetization dynamics, the (fixed) current density generates an Ohmic dissipation power
\begin{align}
 P_J=(j_c wd_n)\frac{l}{\sigma wd_n}=j_c^2\frac{V_n}{\sigma}, \label{seq:Pj-Pt}
\end{align}
where $d_n$ and $\sigma$ are the thickness and conductivity and $V_n=wld_n$ is the volume of the Pt layer. The spin current density $j_s$ delivers spin angular momenta to the YIG by a damping-like torque, which is characterized by the real part of the spin-mixing conductance $G_r$. Similar to what we have done above for the voltage-induced torque, we assume a simple cubic lattice for the whole system. Let $j_c$ be applied along $x$, then the spin polarization is along $y$, and hence the effective SOT field is along $x$ again, which makes it possible to use the same susceptibility $\tilde{\chi}_\parallel$ and $\tilde{\chi}_{\perp}$ derived in Eqs.~\eqref{eq:chix} and~\eqref{eq:chiy}. Using the convention in Ref.~\cite{Ran_PRL_2016} to account for spin diffusion and spin backflow, Eq.~\eqref{seq:pm} gives rise to
\begin{align}
    P_M=\alpha\theta_s^2j_c^2\xi^2\frac{\hbar wla_0^3}{e^2Sd_m}\omega^2\left(|\tilde{\chi}_\perp|^2+|\tilde{\chi}_\parallel|^2\right), \label{seq:Pm-YIG}
\end{align}
where $\theta_s=j_s/j_c$ is the spin Hall angle, $d_m$ is the thickness of YIG, and
\begin{align}
    \xi=\frac{\lambda G_r {\rm tanh}\frac{d_n}{2\lambda}}{\sigma+2\lambda G_r {\rm coth}\frac{d_n}{2\lambda}}
\end{align}
is a renormalization factor due to spin diffusion and spin backflow with $\lambda$ the spin diffusion length of Pt. The mechanical efficiency is then 
\begin{align}
    \eta=\frac{P_M}{P_J}=\frac{\alpha\sigma\hbar\theta_s^2\xi^2}{Se^2}\left(\frac{a_0^3}{d_nd_m}\right)\omega^2\left(|\tilde{\chi}_\perp|^2+|\tilde{\chi}_\parallel|^2\right).
\end{align}
As a typical estimate, $G_r\sim10^{18}$ $\rm m^{-1}$ for YIG/Pt. For $d_n\sim\lambda$, $\xi$ is on the order of $1\%$. At the resonance point, regardless of $\omega_r$ (whose value is much lower in ferromagnetic resonances), $\omega^2\left(|\tilde{\chi}_\perp|^2+|\tilde{\chi}_\parallel|^2\right)$ is roughly $1/\alpha$. Therefore, $\eta$ turns out to be at most $1\%$. If a metallic FM is used, the shunting effect will further reduce the mechanical efficiency.

\section{Effective circuit}\label{sec:appendix circuit}

We seek for an effective circuit to better reflect the electrical response of the considered FM-TI-FM trilayer. To this end, we notice that the electrical impedance of our system is
\begin{align}
    Z_{\rm sys}=\frac{l}{\tilde{\sigma}_{xx}w}=Z_m\left[\frac{-(1+\alpha^2)\omega^2+\omega_r^2+2i\alpha\omega\omega_r}{i\omega\omega_r-\alpha\omega^2}\right], \label{seq:zsys}
\end{align}
where $Z_m^{-1}=\overline{\Omega}_{k_x m^t_x}^2\zeta w/l$. To find hints for the desired effective circuit, we further decompose $Z_{\rm sys}$ as
\begin{align}
    Z_{\rm sys}=&Z_m\frac{(\alpha\omega-i\omega_r)^2+\omega^2}{\omega(\alpha\omega-i\omega_r)} \notag\\
    =&Z_m\left(\alpha+\frac{\omega_r}{i\omega}\right)+\frac{Z_m}{\left(\alpha+\frac{\omega_r}{i\omega}\right)}, \label{seq:zsys-decomp}
\end{align}
where the first term can be represented by a resistor connecting in series with a capacitor and the second term can be represented by a resistor connecting in parallel with an inductor. Accordingly, we conceive an effective circuit depicted in Fig.~\ref{fig:device}, whose impedance is
\begin{align}
    Z_{\rm eff}=r+\frac{1}{i\omega C}+\frac{i\omega LR}{i\omega L+R}.
\end{align}
By setting $Z_{\rm eff}=Z_{\rm sys}$, we obtain Eq.~\eqref{eq:RrLC}.

\bibliography{reference}

\begin{thebibliography}{53}
\bibitem [{Une()}]{Unequallayers}%
  \BibitemOpen
  \href@noop {} {}\bibinfo {note} {If the two FM have different anisotropy
  $\omega_A^t$ and $\omega_A^b$ and $\omega_E\gg\omega_A$, then
  $\omega_r\approx2\omega_E+(\omega_A^t+\omega_A^b)/2$.}\BibitemShut {Stop}%
\bibitem [{Hal()}]{Hallcurrent}%
  \BibitemOpen
  \href@noop {} {}\bibinfo {note} {Here the Hall current arising from
  $\overleftrightarrow{\Omega}_{kk}$ is not included, because it is irrelevant
  to our discussion. The corresponding quantized Hall conductivity $\sigma_H$
  is frequency independent, which can be easily separated from our
  $\sigma_{xy}(\omega)$.}\BibitemShut {Stop}%
\bibitem [{not()}]{note:eta}%
  \BibitemOpen
  \href@noop {} {}\bibinfo {note} {In a current-driven resonance, $\eta$ varies
  significantly with $\omega$, peaking around the resonance frequency
  $\omega_r$. When the system goes off-resonance, $\eta$ also vanishes.
  Comparatively, $\eta=1$ is independent of frequency (if leakage currents are
  ignored) in our system.}\BibitemShut {Stop}%
\end{thebibliography}

\end{document}